\documentclass[twocolumn,showpacs,amsmath,amssymb,pr,superscriptaddress]{revtex4-1}
\usepackage{graphicx}
\usepackage{epsf}
\usepackage{ulem}
\usepackage{color}
\usepackage[unicode=true,colorlinks=true,urlcolor=blue,citecolor=blue]{hyperref}

\begin{document}

\title{AC and DC Conductivities in an n-GaAs/AlAs Heterostructure with a Wide Quantum Well in the Integer Quantum
Hall Effect Regime}

\author{A. A. Dmitriev}
\affiliation{Ioffe Institute, Russian Academy of Sciences, St. Petersburg, 194021 Russia}
\affiliation{ITMO University, St. Petersburg, 197101 Russia}
\author{I. L. Drichko}
\author{I. Yu. Smirnov}
\affiliation{Ioffe Institute, Russian Academy of Sciences, St. Petersburg, 194021 Russia}
\author{A. K. Bakarov}
\affiliation{Rzhanov Institute of Semiconductor Physics, Siberian Branch, Russian Academy of Sciences, Novosibirsk, 630090 Russia}
\author{A. A. Bykov}
\affiliation{Rzhanov Institute of Semiconductor Physics, Siberian Branch, Russian Academy of Sciences, Novosibirsk, 630090 Russia}
\affiliation{Novosibirsk State University, Novosibirsk, 630090 Russia}

\begin{abstract}
The direct-current (dc) $\sigma_{xx}^{dc}$ and alternating-current (ac) $\sigma_{xx}^{ac}=\sigma_1-i\sigma_2$ conductivities of a wide (46 nm) GaAs quantum well with the bilayer electron density distribution are measured. It is found that the magnetic field dependence of $\sigma_{xx}$ exhibits three sets of oscillations related to the transitions between Landau levels in symmetric and antisymmetric subbands and with the transitions occurring owing to the Zeeman splitting of these subbands. The analysis of the frequency dependence of the ac conductivity and the $\sigma_1 / \sigma_2$ ratio demonstrates that the conductivity at the minima of oscillations is determined by the hopping mechanism.
\end{abstract}


\maketitle

\section{Introduction}
Two-dimensional systems with wide quantum wells are a kind of bilayer electron systems. The double quantum wells represent the most popular type of such systems. The bilayer systems consist of two parallel quantum wells filled with the two-dimensional electron gas and separated by a barrier. The tunnel coupling between the quantum wells gives rise to the electron energy spectrum of the bilayer system, which includes the symmetric (S) and antisymmetric (AS) subbands separated by the band gap $\Delta_{SAS} = E_{AS} - E_S$, where $E_{AS}$ and $E_{S}$ are the energies corresponding to the bottoms of $AS$ and $S$ subbands, respectively.

In a usual bilayer system, free electrons are located
in two parallel GaAs quantum wells separated by an
AlGaAs layer ~\cite{bib:Boebinger}. In a wide GaAs quantum well, the
formation of the bilayer system occurs owing to the
Coulomb repulsion of electrons, which pushes them to
the side boundaries of the heterostructure~\cite{bib:Suen}. Such
bilayer system exhibits a higher electron mobility in
comparison to the system based on the double quantum
well. This occurs because the wide GaAs quantum
well does not contain the AlGaAs spacer. In this
case, scattering related to dopants is suppressed and
scattering at the heterostructure boundaries is also
reduced since the number of such boundaries in a single
wide quantum well is half as many as that in the
double well.

The experiments on the integer and fractional Hall
effect ~\cite{bib:Boebinger,bib:Suen,bib:Suen2} stimulated the preparation of bilayer systems
with a high electron mobility and studies of their
transport characteristics. These studies revealed the
role of single-particle and collective effects in the formation
of quantum Hall states. In particular, it was
found that the quantum states for the filling factors $\nu=1$
and 3 in bilayer electron systems can collapse~\cite{bib:Boebinger,bib:Suen}. In addition, it was shown \cite{bib:Shabani} that the bilayer systems
can exhibit the fractional quantum Hall effect
with the occupation numbers having even denominators.

The two-subband electron spectrum manifests itself not only in the quantum Hall effect regime, but also in the case of overlapping Landau levels. In this case, the elastic intersubband scattering of electrons in the two- subband system gives rise to the so-called magneto-intersubband oscillations of the resistivity ~\cite{bib:Polyanovsky,bib:Coleridge,bib:Leadley} in addition to the Shubnikov–de Haas (SdH) oscillations. The period of magneto-intersubband oscillations is
determined by the condition $\Delta_{SAS}=k\hbar \omega_c$, where
$k$=1, 2, 3 …, $\omega_c = eB/m^*$ is the cyclotron frequency, $B$ is the applied magnetic field, and $m^*$ is the effective mass. Magneto-intersubband oscillations in contrast to SdH oscillations are not suppressed by the thermal broadening of the Fermi distribution function. For this reason, they are widely used in the studies of the quantum transport under conditions of suppressed SdH oscillations~\cite{bib:Bykov,bib:Bykov2}.

To study transport phenomena, measurements of the dc and ac conductivities are widely used. These techniques do not exclude but rather supplement each other, and their simultaneous usage provides an opportunity to obtain a more comprehensive information on the transport characteristics of the system under study. Quite recently, the magneto-intersubband oscillations of the ac conductivity have been observed in the bilayer electron system~\cite{bib:Drichko}.

Magneto-intersubband oscillations in bilayer systems were studied in many works, whereas the specific features of SdH oscillations and the integer quantum Hall effect remain almost unstudied. We are aware of only one work ~\cite{bib:Boebinger2} where these features were observed, but not analyzed.

Our study is aimed at revealing the role of the two- subband spectrum for the dc and ac conductivities in the high-mobility bilayer electron system in quantizing magnetic fields.

\section{Samples and Methods}

\begin{figure}[t]
\centerline{
\includegraphics[width=8.2cm,clip=]{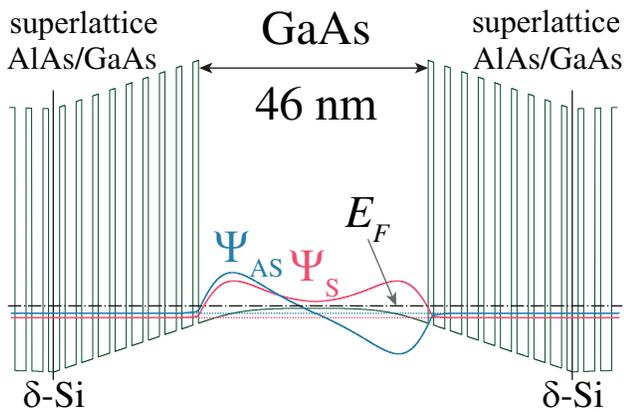}}
\caption{ Fig.~1. Layout of the sample.} \label{sample}
\end{figure}

The sample under study is a symmetric n-doped wide (46 nm) GaAs quantum well. AlAs/GaAs superlattices are used as the barriers. The layout of the sample is shown in Fig.~\ref{sample}. The sample was grown by the molecular-beam epitaxy technique at the (100) GaAs substrate.

In the studies of the conductivity in the temperature range of 1.7-4.2 K, we use three methods: the dc measurements and the contactless acoustic and microwave spectroscopy methods.

The $\rho_{xx}$ and $\rho_{xy}$ components of the dc resistivity are measured using Hall bars 450 $\mu$m long and 50 $\mu$m wide. The electric current density does not exceed 200 $\mu$A/cm. The measurements are performed at temperatures from 2.2 to 4.2 K at the applied electric field up to 14 T. The analysis of the dc resistivity allows determining the filling factors for the Landau levels on the basis of the values corresponding to the quantum Hall effect plateau at high magnetic fields.

The study of the frequency and temperature dependences of the ac conductivity makes it possible to reveal the mechanisms of the low-temperature conductivity. For this purpose, we use two contactless techniques. One of them is the acoustic technique involving a Rayleigh surface acoustic wave (SAW) propagating along the surface of a LiNbO$_3$ piezoelectric substrate, against which the sample under study is pressed by a spring. The strain is not transferred to the sample. Owing to the piezoelectric effect, this wave is accompanied by the electric field wave with the same frequency. Such field penetrates into the sample and interacts with charge carriers in the conduction channel. This leads to the absorption of the SAW and changes its velocity. Note that the polarization vector of the electric field is parallel to the wave vector of the SAW. The absorption coefficient $\Gamma$ (in decibels per centimeter) and changes in the SAW velocity $\Delta V/V_0$ as functions of the ac conductivity components $\sigma_{xx}^{ac} = \sigma_1 - i \sigma_2$ are determined by the expressions
derived in ~\cite{bib:Kagan,bib:Drichko2}:
\begin{eqnarray}
  \label{eq:G}
&&\Gamma =8.68\frac{K^2}{2}qA   \frac{\Sigma_1}
  {[1+\Sigma_2]^2+[\Sigma_1]^2},  \frac{\text{dB}}{\text{cm}} \,  \nonumber  \\
&&\frac{\Delta V}{V_0}=\frac{K^2}{2}A\frac{1+\Sigma_2}
  {[1+\Sigma_2]^2+[\Sigma_1]^2} \,  \\
&&\Sigma_i=4 \pi \sigma_i t(q)/\varepsilon_s V_0 \,  \nonumber  \\
&&A = 8b(q)(\varepsilon_1 +\varepsilon_0)
\varepsilon_0^2 \varepsilon_s
\exp [-2q(a+d)], \nonumber
\end{eqnarray}
Here, $K^2$ is the electromechanical coupling coefficient of lithium niobate; $q$ and $V_0$ are the wave vector
and velocity of the SAW in LiNbO$_3$, respectively; $a$ is
the spacing between the piezoelectric lithium niobate substrate and the sample; $d$ is the depth at which the conduction channel is located (it is specified by technologists); $\varepsilon_1$,
$\varepsilon_0$ and $\varepsilon_s$ are the permittivities of lithium niobate, vacuum, and the sample, respectively; and b and t are complicated functions depending on $a$, $d$, $\varepsilon_1$,
$\varepsilon_0$ and $\varepsilon_s$. A solution of this set of equations makes it possible to determine $\sigma_1$ and $\sigma_2$.

The advantage of such technique is the possibility of determining simultaneously the real, $\sigma_1$, and imaginary, $\sigma_2$, components of the ac conductivity. However, it allows measurements only at some definite frequencies corresponding to the odd harmonics of the interdigital transducers used to excite and detect SAWs. In our case, the allowed frequencies are 30, 90, 150, and 196 MHz.

In the analysis of the conductivity as a function of the SAW frequency, it is very important to perform measurements in a wide frequency range. Such a possibility is provided by the microwave technique. In this technique, a high-frequency electric field is applied to a two-dimensional electron system through a coplanar waveguide, onto which the sample is mounted. The waveguide is manufactured at the insulating i-GaAs substrate in the shape of a meander to increase the length of coupling with the two-dimensional electron system. Similar to the acoustic technique, the interaction of the electric field of the propagating quasi-TEM wave with charge carriers in the conduction channel of the sample reduces its intensity and leads to a phase shift. In the microwave technique, the working frequency range is wider and is in fact determined by the characteristics of the used measurement equipment (in our case, the frequency ranges from 150 to  1200 MHz) rather than by the configuration of the coplanar waveguide. In this case, the conductivity is given by the expression~\cite{bib:Engel}:
\begin{eqnarray}
\sigma_1=-\frac{w}{Z_0 l} \ln \left[\frac{U_{out}}{U_{in}}\right]
\sqrt{1+\left[\frac{v_{ph}}{l \omega}
\ln\left(\frac{U_{out}}{U_{in}}\right)\right]^2},
\label{s1CPW}
\end{eqnarray}
Here, $U_{in}$ and $U_{out}$ are the amplitudes of the input and output signals, respectively; $Z_0$=50~$\Omega$ is the characteristic impedance of the coplanar waveguide without the sample; $l$=5.3 cm is the length of the signal wire for the coplanar waveguide; $w$=26 $\mu$m is the width of the
gap between the signal and ground wires; $v_{ph} = c
\sqrt{2/(1+\varepsilon_{\text{GaAs}})}$=1.14$\times$10$^{10}$
cm/s is phase velocity of the wave  propagating along the waveguide; and
$\varepsilon_{\text{GaAs}}$=12.9 is the relative permittivity of the i-GaAs substrate, at which the coplanar waveguide is formed.

According to Eq.~\ref{s1CPW}, to determine $\sigma_1$, it is necessary to know the amplitude of the input signal, which is quite difficult to measure accurately. In addition, to
obtain the exact value of $\sigma_1$, it is necessary to ensure the complete matching of the system, i.e., the absence of reflections at the edges of the coplanar waveguide.

To obtain absolute values of $\sigma_1$ measured using this technique, we employ the calibration procedure
described in detail in~\cite{bib:ourCPW}. This procedure involves the assumption that the conductivity at the peaks of SdH oscillations and of the integer quantum Hall effect occurs via delocalized states and is independent of the frequency in the range under study.

The ac conductivity was measured by both of these techniques at temperatures from 1.7 to 4.2 K and at magnetic fields up to 8 T.

\section{Experiment}
\subsection{Direct-Current Measurement Results}

\begin{figure}[t]
\centerline{
\includegraphics[width=8.2cm,clip=]{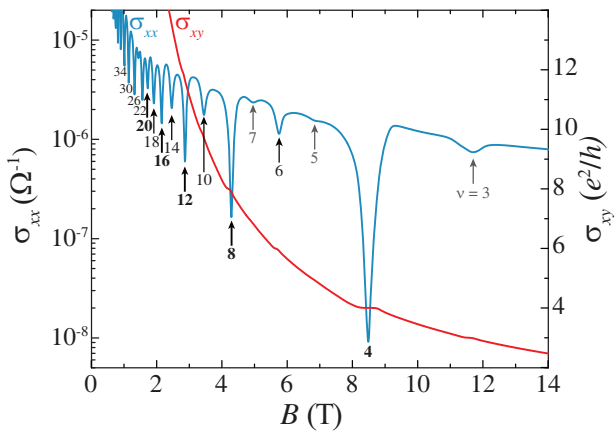}}
\caption{Magnetic field dependence of
$\sigma_{xx}$ and  $\sigma_{xy}$ at 2.2~K. The arrows indicate the filling factors.} \label{sxxxyDC}
\end{figure}

In Fig.~\ref{sxxxyDC}, we show the magnetic field dependence
of the diagonal, $\sigma_{xx}$, and Hall, $\sigma_{xy}$, conductivities at
$T$=2.2 K. Using measured $\rho_{xx}$ and $\rho_{xy}$ values, these components are calculated by the standard formulas
\begin{eqnarray}
\sigma_{xx}=\frac{\rho_{xx}}{\rho_{xx}^2+\rho_{xy}^2},
\text{   }
\sigma_{xy}=\frac{\rho_{xy}}{\rho_{xx}^2+\rho_{xy}^2}.
 \label{sDC}
\end{eqnarray}

The Hall conductivity exhibits a plateau corresponding
to the integer quantum Hall effect, whereas
the diagonal conductivity is characterized by a rich
picture of oscillations. At magnetic fields up to 3.5 T,
we observe SdH oscillations, which correspond to the
regime of the quantum Hall effect.

The Hall conductivity at the quantum Hall effect
plateau is determined by the filling factor $\nu$:
\begin{eqnarray}
\sigma_{xy}=\nu\frac{e^2}{h}.
 \nonumber 
\end{eqnarray}

The filling factors $\nu$=3, 4, 5, 6, and 8 are determined
from the plateaus observed at $T$=2.2~K. The total electron density $n_t$=8.27$\times$10$^{11}$~cm$^{-2}$ is determined from the slope of the linear $\nu (1/B)$ dependence.

The filling factors assigned to the oscillations are indicated in Fig.~\ref{sxxxyDC}. Here, an unusual situation occurs: we can distinguish three rather than two (as usual) sets of oscillations differing in their amplitudes: (i) deep oscillations, $\nu=4N$, where $N$ is an integer; (ii) oscillations of a lower depth, $\nu=4N$+2; and (iii) very weak oscillations with odd $\nu$ values. The nature of this phenomenon is discussed below in the next section.

\subsection{Acoustic and Microwave Spectroscopy Results}
In Fig. 3, we show the magnetic field dependence of the absorption coefficient $\Gamma$ and relative change $\Delta V/V_0$ of the SAW velocity measured by the acoustic technique at 1.7 K. In Fig.~\ref{s12}(a), we illustrate the magnetic field dependence of the real and imaginary components of the ac conductivity calculated by Eqs.\ref{eq:G}. It is seen that $\sigma_2 > \sigma_1$ at magnetic fields $B$=2.8~T, 4.2~T and 5.7~T, at which the minima of the electrical conductivity are observed, whereas $\sigma_2 < \sigma_1$ in the magnetic field ranging between the minima and at low amplitude oscillations. The magnetic field dependence of $\sigma_2 / \sigma_1$ is presented in Fig.~\ref{s12}(b).

\begin{figure}[t]
\centerline{
\includegraphics[width=8.2cm,clip=]{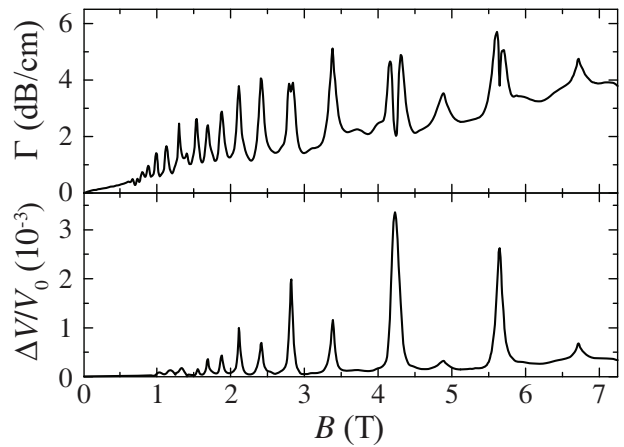}}
\caption{Magnetic field dependence of the absorption coefficient $\Gamma$ and the relative change in the velocity $\Delta V/V_0$ at
 $T$=1.7~K, $f$=30~MHz} \label{GV}
\end{figure}

\begin{figure}[t]
\centerline{
\includegraphics[width=8.2cm,clip=]{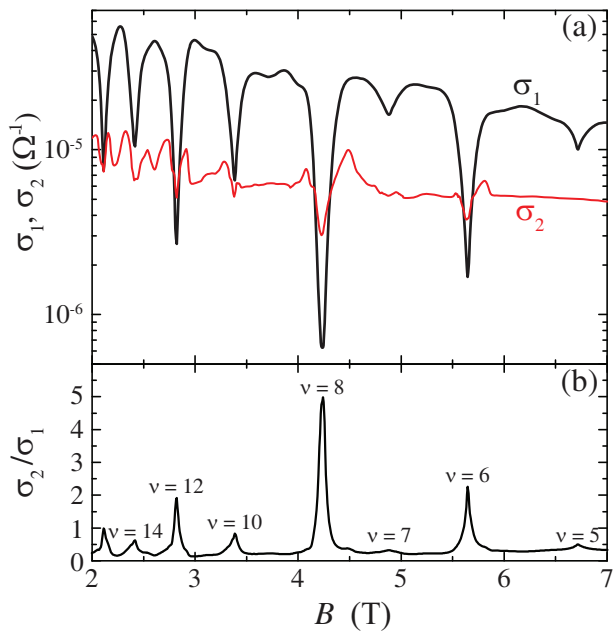}}
\caption{(a) Magnetic field dependence of the real ($\sigma_1$) and imaginary ($\sigma_2$) parts of the ac conductivity at $T$=1.7~K. (b) Magnetic field dependence of $\sigma_2 / \sigma_1$.} \label{s12}
\end{figure}

\begin{figure}[t]
\centerline{
\includegraphics[width=8.2cm,clip=]{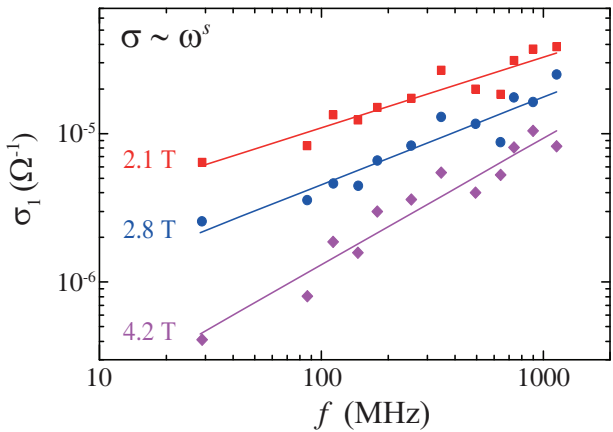}}
\caption{Frequency dependence of $\sigma_1$ at applied magnetic fields of 2.1, 2.8, and 4.2 T and $T$=1.7~K.} \label{s1f}
\end{figure}

In Fig.~\ref{s1f}, we show the frequency dependence of the conductivity at the minima of oscillations corresponding to different filling factors. From Fig.~\ref{s1f}, it is clear that the frequency dependence of the conductivity can
be fitted by the power law $\sigma \propto \omega^s$.

\section{Discussion}
For the interpretation of three sets of SdH oscillations and the integer quantum Hall effect, we draw the energy diagram of the quasi-two-dimensional electron system under study at the applied magnetic field. In this system with the wide quantum well, the Coulomb repulsion of electrons gives rise to the formation of $S$ and $AS$ subbands separated by the energy gap $\Delta_{SAS}$. In the applied magnetic field, each of these subbands generates a “staircase” of Landau levels. To construct the energy diagram, it is necessary to determine the energy gap $\Delta_{SAS}$. The Fourier analysis of the magnetoresistance at magnetic fields $B <$ 1~T gives
$\Delta_{SAS}$1.5 meV. The total electron density $n_t$ is determined
from the dc measurements. The knowledge of the total electron density and $\Delta_{SAS}$ makes it possible to determine the electron density in the symmetric and antisymmetric subbands, which appear to be $n^{(1)}$=4.27$\times$10$^{11}$~cm$^{-2}$ and $n^{(2)}$=3.93$\times$10$^{11}$~cm$^{-2}$, respectively.

\begin{figure*}[t]
\centerline{
\includegraphics[width=14 cm,clip=]{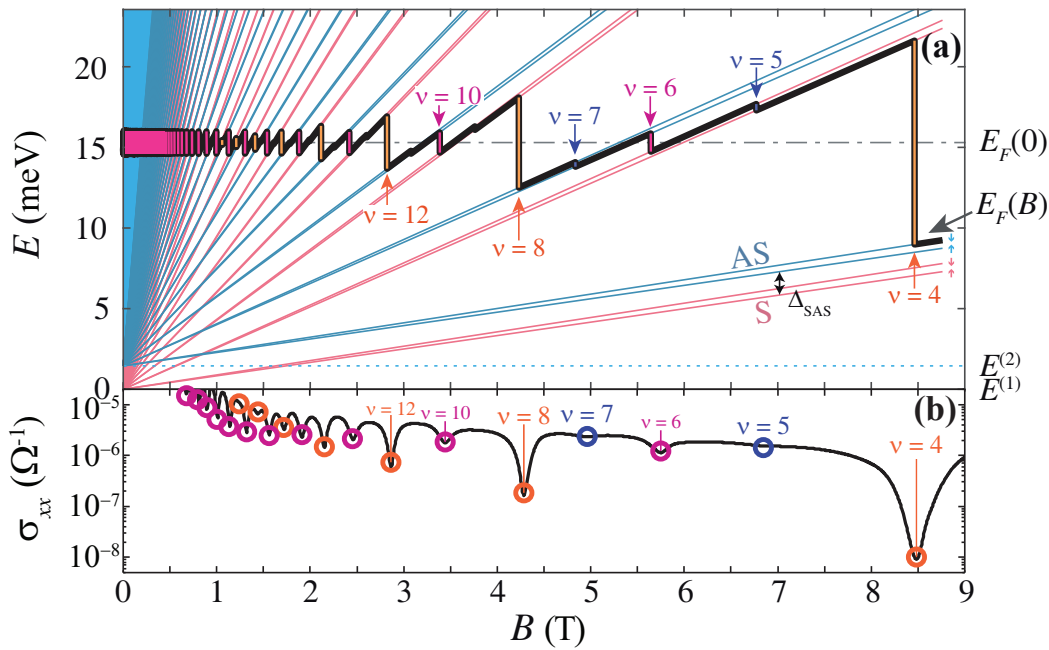}}
\caption{(a) Energy diagram (the “fan” of Landau levels) for the quasi-two-dimensional electron structure in the applied magnetic field. Thin solid lines denote the energies of Landau levels. The dash-dotted line corresponds to the Fermi energy at zero magnetic field, whereas the thick solid line illustrates the magnetic field dependence of the Fermi energy. (b) Magnetic field dependence of the conductivity $\sigma^{dc}_{xx}$.}
\label{LLfan}
\end{figure*}

The energy diagram is shown in Fig.~\ref{LLfan}. The energy is measured from the bottom of the $S$ subband. Thus, at zero magnetic field, the position of the Fermi level shown in Fig.~\ref{LLfan} by the dash-dotted line appears to be at 15.2 meV. To explain the position and amplitude of magnetic field induced oscillations, we calculate and plot the magnetic field dependence of the Fermi energy (the solid line in Fig.~\ref{LLfan} (a)). In Fig.~\ref{LLfan} (b), we show for comparison the magnetic field dependence of the conductivity. It is seen that the positions of the Fermi level jumps related to the transitions between the Landau levels illustrated in Fig.~\ref{LLfan} (a) indeed correspond to the magnetic field positions of the minima of conductivity oscillations. These transitions can be conditionally classified into four types: (i) the transitions between the spin-split energies in the S subband, (ii)	the transitions from the $N$th Landau level in the $S$ subband to the $N$th Landau level in the $AS$ subband,
(iii)	the transitions between the spin-split energies in the $AS$ subband, and (iv) the transitions from the ($N$-1)th Landau level in the AS subband to the $N$th Landau level in the S subband. In the activation conductivity range, the deep oscillations with $\nu=4N$
correspond to the transitions of type (iv), the oscillations of a lower depth with $\nu=4N$+2 correspond to the transitions of type (ii), and the very weak oscillations with odd $\nu$ values correspond to the transitions of types (i) and (iii).

The transitions of types (i) and (iii) corresponding to the oscillations with $\nu=4N$+1 and $\nu=4N$+3 are due to the spin splitting of Landau levels, for which the activation energy is determined by the Zeeman energy. This energy is low; hence, the depth of these oscillations is also low.

For the transitions of type (ii) related to the oscillations with $\nu=4N$+2, the corresponding activation energy disregarding the collisional broadening is $\Delta_{SAS}-\Delta_Z$. Under the condition $\Delta_{SAS} \gg \Delta_Z$, the activation energy is nearly independent of the magnetic field. This explains the absence of magnetic field dependence for the amplitude of oscillations related to these transitions beginning with low (from 0.5 T) magnetic field values.

For the transitions of type (iv) ($\nu=4N$), the corresponding activation energy disregarding the broadening of Landau levels is  $\hbar \omega_c-\Delta_{SAS}-\Delta_Z$.  As shown above, the Zeeman splitting $\Delta_Z$ is negligibly small compared to $\hbar \omega_c$. In the magnetic field $B \approx$1~T,
$B \approx$1~T $\hbar \omega_c = \Delta_{SAS}$, and the activation energy vanishes; i.e., the oscillations with $\nu=4N$ should disappear. This is indeed observed in experiment at the fields below 1 T, at which only the oscillations with $\nu=4N$+2 (see Fig.~\ref{sxxxyDC}). At magnetic fields $B <$2~T $\hbar \omega_c-\Delta_{SAS} < \Delta_{SAS}$, and the oscillations with $\nu=4N$ should
have a depth lower than that for the oscillations with
$\nu=4N$+2. This is also observed in experiment (see Fig.~\ref{sxxxyDC}). At magnetic fields $B <$2~T, the conductivity activation energy at the minima for the oscillations with $\nu=4N$ is the highest, and the oscillations are the deepest.

Now, we discuss the mechanisms of conduction. At the peaks of oscillations, the ac conductivity is independent of the frequency, which is characteristic of the conduction via delocalized states~\cite{bib:Drichko2}. At the minima of oscillations, the charge carriers are localized, and the ac hopping conductivity occurs. It is convenient to describe such conductivity using the two-site model. If $\omega \tau \ll$1 (where $\omega$ is the SAW frequency and $\tau$ is the characteristic time for the changes in the occupation of the levels in the two-level system in the electric field
of the SAW), then $\sigma_1 \propto \omega^s$, where $s \approx$1, and $\sigma_2 > \sigma_1$~\cite{bib:Efros,bib:Parshin}.

Indeed, this situation is illustrated in Fig.~\ref{s1f}, where we show the dependence of the ac conductivity at the oscillation minima on the SAW frequency at different applied magnetic fields. It is seen that this dependence
can be fitted by the power law $\sigma_1 \propto \omega^s$ with the exponent $s$ ranging from 0 to 1 depending on the depth of
oscillations. The larger the exponent s, the lower the conductivity at the oscillation minimum.

The magnetic field dependence of $\sigma_2 / \sigma_1$ is shown in Fig.\ref{s12}(b). In Fig.~\ref{s12}(b), we can see that $\sigma_2 / \sigma_1$ grows with the amplitude of oscillations. Indeed, the deeper the oscillation, the lower the conductivity at the minimum, and the larger the number of charge carriers that appear to be in localized states. In this case, the frequency dependence of the ac conductivity tends to the
law $\sigma_1 \propto \omega^1$, whereas $\sigma_2 / \sigma_1$ increases and far exceeds unity.

\section{Conclusion} The effect of the two-subband spectrum on the electron conductivity has been studied in the integer quantum Hall regime for the heterostructure with the wide (46 nm) n-GaAs quantum well. The energy diagram for the two-subband system has been constructed. This diagram allows explaining three sets of oscillations observed in the experiment. At the peaks of oscillations, the ac conductivity is independent of the frequency, which is characteristic of the conductivity via delocalized states. At the oscillation minima, the charge carriers are localized, and the hopping conductivity occurs. This is supported by the observed frequency dependence of the conductivity, which can be fitted by a power relation with the exponent increasing with the degree of localization. The degree of localization is characterized by the relation between the real
and imaginary part of the ac conductivity, $\sigma_1 < \sigma_2$.

\section{Acknowledgments}
This work was partially supported by the Presidium of the Russian Academy of Sciences and by the Russian Foundation for Basic Research, project no. 19-02-00124.

\vfill\eject

\end{document}